\documentclass[aps,twocolumn,showpacs,preprintnumbers,amsmath,amssymb,
nofootinbib,showkeys,floatfix]{revtex4-1}

\usepackage{latexsym}
\usepackage{amsmath}
\usepackage{amssymb}

\usepackage{supertabular} 
\usepackage{placeins}
\usepackage{epsfig}
\usepackage{graphicx}
\usepackage{graphics}

\def\tstrut{\vrule height2.25ex depth0pt width0pt} 

\usepackage{hyperref} 
\hypersetup{
   colorlinks=true,
   linkcolor=blue,
   citecolor=blue,
   urlcolor=blue
}

\bibpunct{[}{]}{,}{n}{,}{,} 

\begin{document}

\title{Charmed-strange Meson Spectrum: Old and New Problems}
\author{Jorge Segovia}\email{segonza@usal.es}\affiliation{Grupo de F\'isica
Nuclear and Instituto Universitario de F\'isica Fundamental y Matem\'aticas
(IUFFyM) \\ Universidad de Salamanca, E-37008 Salamanca, Spain}
\author{David R. Entem}\email{entem@usal.es}\affiliation{Grupo de F\'isica
Nuclear and Instituto Universitario de F\'isica Fundamental y Matem\'aticas
(IUFFyM) \\ Universidad de Salamanca, E-37008 Salamanca, Spain}
\author{Francisco Fern\'andez}\email{fdz@usal.es}\affiliation{Grupo de F\'isica
Nuclear and Instituto Universitario de F\'isica Fundamental y Matem\'aticas
(IUFFyM) \\ Universidad de Salamanca, E-37008 Salamanca, Spain}
\date{\today}

\begin{abstract}
The LHCb Collaboration has recently reported the observation for the first time
of a spin-$3$ resonance in the heavy quark sector. They have shown that the
$\bar{D}^{0}K^{-}$ structure seen in the $B_{s}^{0}\to\bar{D}^{0}K^{-}\pi^{+}$
reaction and with invariant mass $2.86\,{\rm GeV}$ is an admixture of a spin-$1$
and a spin-$3$ resonances. Motivated by the good agreement between our
theoretical predictions some time ago and the properties extracted from the
experiment of the $D_{s1}^{\ast}(2860)$ and $D_{s3}^{\ast}(2860)$ states, we
perform an extension of the study of the strong decay properties of the
$D_{sJ}^{\ast}(2860)$ and present the same analysis for the
$D_{s1}^{\ast}(2700)$ and $D_{sJ}(3040)$ mesons. This provides a unified and
simultaneous description of the three higher excited charmed-strange resonances
observed until now. For completeness, we present theoretical results for masses
and strong decays of the low-lying charmed-strange mesons and those experimental
missing states which belong to the spin-multiplets of the discovered
$D_{s1}^{\ast}(2700)$, $D_{sJ}^{\ast}(2860)$ and $D_{sJ}(3040)$ resonances. The
theoretical framework used is a constituent quark model which successfully
describes hadron phenomenology from light to heavy quark sectors.
\end{abstract}


\maketitle


\section{INTRODUCTION}
\label{sec:introduction}

The spectrum of charmed-strange mesons contains a number of well established
states~\cite{Beringer:1900zz} corresponding to the $S$-wave $D_{s}$ and
$D_{s}^{\ast}$ mesons with spin-parity $0^{-}$ and $1^{-}$, respectively; and
the $P$-wave states with quantum numbers $J^{P}=0^{+}$ ($D_{s0}^{\ast}(2317)$),
$1^{+}$ ($D_{s1}(2460)$ and $D_{s1}(2536)$) and $2^{+}$ ($D_{s2}^{\ast}(2573)$).

In addition, between the years $2006$ and $2009$, three new $c\bar{s}$ mesons
were observed at the $B$-factories in $DK$ and $D^{\ast}K$ decay modes and in
three-body $b$-hadron
decays~\cite{Aubert:2006mh,Brodzicka:2007aa,Aubert:2009ah}. These states have
been recently included in the Particle Data Group (PDG) as the
$D_{s1}^{\ast}(2700)$, the $D_{sJ}^{\ast}(2860)$ and the $D_{sJ}(3040)$. While
the $D_{s1}^{\ast}(2700)$ is commonly believed to have quantum numbers
$J^{P}=1^{-}$, there are several possibilities for the $D_{sJ}^{\ast}(2860)$
and $D_{sJ}(3040)$ states. Different predictions of the theoretical models
can be found in
Refs.~\cite{Swanson:2006st,Rosner:2006jz,Klempt:2007cp,Colangelo:2012xi}.

Recent experiments performed by the LHCb Collaboration have contributed to
clarify the puzzle around the $D_{sJ}^{\ast}(2860)$
resonance~\cite{Aaij:2014xza,Aaij:2014baa}. A careful re-examination of the
$\bar{D}^{0}K^{-}$ invariant mass around $2.86\,{\rm GeV}$ in the decay
$B_{s}^{0}\to\bar{D}^{0}K^{-}\pi^{+}$ finds that a spin-$1$ state and a spin-$3$
state overlap under the peak. Since the resonance substructure of the three-body
decay is analysed through a Dalitz plot, the well-defined initial state allows
to unambiguously determine the quantum numbers, in particular, the parity of the
$D_{sJ}^{\ast}(2860)$ state to be odd. The determined masses and widths of the
$D_{s1}^{\ast}(2860)$ and $D_{s3}^{\ast}(2860)$
are~\cite{Aaij:2014xza,Aaij:2014baa}
\begin{equation}
\begin{split}
M(D_{s1}^{\ast}(2860)) &= (2859 \pm 12 \pm 6 \pm 23) \, {\rm MeV}, \\ 
\Gamma(D_{s1}^{\ast}(2860)) &= (159 \pm 23 \pm 27 \pm 72) \, {\rm MeV},
\label{eq:ExpInfDs12860} 
\end{split}
\end{equation}
and
\begin{equation}
\begin{split}
M(D_{s3}^{\ast}(2860)) &= (2860.5 \pm 2.6 \pm 2.5 \pm 6.0) \, {\rm MeV}, \\
\Gamma(D_{s3}^{\ast}(2860)) &= (53 \pm 7 \pm 4 \pm 6) \, {\rm MeV},
\label{eq:ExpInfDs32860}
\end{split}
\end{equation}
where the first uncertainty is statistical, the second is systematic and the
third is due to model variations.

Spin-$3$ states had never been observed in heavy flavoured mesons and so these
new measurements have led to many theoretical
works~\cite{Song:2014mha,Wang:2014jua,Godfrey:2014fga,Zhou:2014ytp, Ke:2014ega,
Song:2015nia}. In Ref.~\cite{Segovia:2012cd} we predicted two resonances
$1^{3}D_{1}$ and $1^{3}D_{3}$ at $2.86\,{\rm GeV}$ with total decay
widths $153\,{\rm MeV}$ and $85\,{\rm MeV}$, respectively\footnote{Note that we
are using here spectroscopic notation: $n^{2S+1}L_{J}$, where $n$ refers to the
radial excitation with $n=1$ indicating the ground state, and $S$, $L$ and $J$
is the spin-, angular- and total-momentum of the $c\bar{s}$ pair.}.

Motivated by the excellent agreement with the LHCb results, we will address in
this work an extension of our study to all strong decay properties of the
$D_{sJ}^{\ast}(2860)$ resonance. Moreover, we will present the same analysis for
the $D_{s1}^{\ast}(2700)$ and $D_{sJ}(3040)$ mesons. We will work within the
framework of a constituent quark model (CQM) proposed in
Ref.~\cite{Vijande:2004he} (see references~\cite{Valcarce:2005em}
and~\cite{Segovia:2013wma} for reviews). This model successfully describes
hadron phenomenology and hadronic
reactions~\cite{Fernandez:1992xs,Garcilazo:2001md,Vijande:2004at} and has
recently been applied to mesons containing heavy quarks (see, for instance,
Refs.~\cite{Segovia:2008zz,Segovia:2009zz,Segovia:2011zza,Segovia:2014mca}).

For completeness, we will resume our theoretical results for the masses of
the low-lying charmed-strange mesons and calculate strong decays for those
states above the open-flavour threshold, the $D_{s1}(2536)$ and
$D_{s2}^{\ast}(2573)$ mesons. There are $c\bar{s}$ states which have not yet
been seen by experiments but belong to the spin-multiplets of the discovered
$D_{s1}^{\ast}(2700)$, $D_{sJ}^{\ast}(2860)$ and $D_{sJ}(3040)$ resonances. We
will also compute their masses and strong decay properties in order to guide
experimentalists in their search.

This manuscript is arranged as follows. In Sec.~\ref{sec:theoreticalframework}
we will describe the main properties of the constituent quark model relevant
to the heavy quark sector and review the $^{3}P_{0}$ strong decay model adapted
to our formalism. In Sec.~\ref{sec:results} we will present our
theoretical results. First, we will review our quark model results for the
properties of the low-lying charmed-strange mesons; second, we will compare the
available experimental data of the $D_{s1}^{\ast}(2700)$, $D_{sJ}^{\ast}(2860)$
and $D_{sJ}^{\ast}(3040)$ resonances with the theoretical predictions attending
to our quantum number assignments. And third, we will compute masses and strong
decay properties of those states that are experimentally missing and lie in the
same energy range of the $D_{s1}^{\ast}(2700)$, $D_{sJ}^{\ast}(2860)$ and
$D_{sJ}^{\ast}(3040)$ resonances. Partial decay widths into all open-decay
channels will be provided in order to guide experimentalists in the quest of
completing the information of the charmed-strange meson sector. We will
summarize and give some conclusions in Sec.~\ref{sec:conclusions}.


\section{THEORETICAL FRAMEWORK}
\label{sec:theoreticalframework}

\subsection{The constituent quark model}
\label{subsec:CQM}

Spontaneous chiral symmetry breaking of the QCD Lagrangian together with the
perturbative one-gluon exchange (OGE) and the nonperturbative confining
interaction are the main pieces of constituent quark models. Using this idea,
Vijande {\it et al.}~\cite{Vijande:2004he} developed a model of the quark-quark
interaction which is able to describe meson phenomenology from the light to the
heavy quark sector.

The wide energy range needed to provide a consistent description of light,
strange and heavy mesons requires an effective scale-dependent strong coupling
constant. We use the frozen coupling constant~\cite{Vijande:2004he}
\begin{equation}
\alpha_{s}(\mu)=\frac{\alpha_{0}}{\ln\left(
\frac{\mu^{2}+\mu_{0}^{2}}{\Lambda_{0 }^{2}} \right)},
\end{equation}
in which $\mu$ is the reduced mass of the $q\bar{q}$ pair and $\alpha_{0}$,
$\mu_{0}$ and $\Lambda_{0}$ are parameters of the model determined by a global
fit to the meson spectra.

In the heavy quark sector chiral symmetry is explicitly broken and
Goldstone-boson exchanges do not appear. Thus, OGE and confinement are the only
interactions remaining. The one-gluon exchange potential contains central,
tensor and spin-orbit contributions given by
\begin{widetext}
\begin{equation}
\begin{split}
&
V_{\rm OGE}^{\rm C}(\vec{r}_{ij}) =
\frac{1}{4}\alpha_{s}(\vec{\lambda}_{i}^{c}\cdot
\vec{\lambda}_{j}^{c})\left[ \frac{1}{r_{ij}}-\frac{1}{6m_{i}m_{j}} 
(\vec{\sigma}_{i}\cdot\vec{\sigma}_{j}) 
\frac{e^{-r_{ij}/r_{0}(\mu)}}{r_{ij}r_{0}^{2}(\mu)}\right], \\
& 
V_{\rm OGE}^{\rm T}(\vec{r}_{ij})=-\frac{1}{16}\frac{\alpha_{s}}{m_{i}m_{j}}
(\vec{\lambda}_{i}^{c}\cdot\vec{\lambda}_{j}^{c})\left[ 
\frac{1}{r_{ij}^{3}}-\frac{e^{-r_{ij}/r_{g}(\mu)}}{r_{ij}}\left( 
\frac{1}{r_{ij}^{2}}+\frac{1}{3r_{g}^{2}(\mu)}+\frac{1}{r_{ij}r_{g}(\mu)}\right)
\right]S_{ij}, \\
&
\begin{split}
V_{\rm OGE}^{\rm SO}(\vec{r}_{ij})= &  
-\frac{1}{16}\frac{\alpha_{s}}{m_{i}^{2}m_{j}^{2}}(\vec{\lambda}_{i}^{c} \cdot
\vec{\lambda}_{j}^{c})\left[\frac{1}{r_{ij}^{3}}-\frac{e^{-r_{ij}/r_{g}(\mu)}}
{r_{ij}^{3}} \left(1+\frac{r_{ij}}{r_{g}(\mu)}\right)\right] \times \\ & \times 
\left[((m_{i}+m_{j})^{2}+2m_{i}m_{j})(\vec{S}_{+}\cdot\vec{L})+
(m_{j}^{2}-m_{i}^{2}) (\vec{S}_{-}\cdot\vec{L}) \right],
\end{split}
\end{split}
\end{equation}
\end{widetext}
where $r_{0}(\mu)=\hat{r}_{0}\frac{\mu_{nn}}{\mu_{ij}}$ and
$r_{g}(\mu)=\hat{r}_{g}\frac{\mu_{nn}}{\mu_{ij}}$ are regulators which depend on
$\mu_{ij}$, the reduced mass of the $q\bar{q}$ pair. The contact term of the
central potential has been regularized as
\begin{equation}
\delta(\vec{r}_{ij})\sim\frac{1}{4\pi
r_{0}^{2}}\frac{e^{-r_{ij}/r_{0}}}{r_{ij}}.
\end{equation}

One characteristic of the model is the use of a screened linear confinement
potential. This has been able to reproduce the degeneracy pattern observed for
the higher excited states of light mesons~\cite{Segovia:2008zza}. As we assume
that confining interaction is flavour independent, we hope that this form of the
potential will be useful in our case because we are focusing on the high energy
region of the charmed-strange meson spectrum.

The different pieces of the confinement potential are
\begin{equation}
\begin{split}
&
V_{\rm CON}^{\rm C}(\vec{r}_{ij})=\left[-a_{c}(1-e^{-\mu_{c}r_{ij}})+\Delta
\right] (\vec{\lambda}_{i}^{c}\cdot\vec{\lambda}_{j}^{c}), \\
&
\begin{split}
&
V_{\rm CON}^{\rm SO}(\vec{r}_{ij}) =
-(\vec{\lambda}_{i}^{c}\cdot\vec{\lambda}_{j}^{c}) \frac{a_{c}\mu_{c}e^{-\mu_{c}
r_{ij}}}{4m_{i}^{2}m_{j}^{2}r_{ij}} \times \\
&
\times \left[((m_{i}^{2}+m_{j}^{2})(1-2a_{s})\right. \\
&
\quad\,\, +4m_{i}m_{j}(1-a_{s}))(\vec{S}_{+} \cdot\vec{L}) \\
&
\left. \quad\,\, +(m_{j}^{2}-m_{i}^{2}) (1-2a_{s}) (\vec{S}_{-}\cdot\vec{L})
\right].
\end{split}
\end{split}
\end{equation}
where $a_{s}$ controls the mixture between the scalar and vector Lorentz
structures of the confinement. At short distances this potential presents a
linear behaviour with an effective confinement strength
$\sigma=-a_{c}\,\mu_{c}\,(\vec{\lambda}^{c}_{i}\cdot \vec{\lambda}^{c}_{j})$,
while it becomes constant at large distances. This type of potential shows a
threshold defined by
\begin{equation}
V_{\rm thr}=\{-a_{c}+ \Delta\}(\vec{\lambda}^{c}_{i}\cdot
\vec{\lambda}^{c}_{j}).
\end{equation}
No $q\bar{q}$ bound states can be found for energies higher than this
threshold.

Table~\ref{tab:parameters} shows the model parameters used herein. Further
details about the quark model and the fine-tuned model parameters can be found
in Refs.~\cite{Vijande:2004he,Segovia:2008zza,Segovia:2008zz}. 

\begin{table}[!t]
\begin{center}
\begin{tabular}{ccc}
\hline
\hline
\tstrut
Quark masses & $m_{n}$ (MeV) & $313$ \\
             & $m_{s}$ (MeV) & $555$ \\
             & $m_{c}$ (MeV) & $1763$ \\
             & $m_{b}$ (MeV) & $5110$ \\
\hline
OGE & $\hat{r}_{0}$ (fm) & $0.181$ \\
    & $\hat{r}_{g}$ (fm) & $0.259$ \\
    & $\alpha_{0}$ & $2.118$ \\
    & $\Lambda_{0}$ $(\mbox{fm}^{-1})$ & $0.113$ \\
    & $\mu_{0}$ (MeV) & $36.976$ \\
\hline
Confinement & $a_{c}$ (MeV) & $507.4$ \\
            & $\mu_{c}$ $(\mbox{fm}^{-1})$ & $0.576$ \\
            & $\Delta$ (MeV) & $184.432$ \\
            & $a_{s}$ & $0.81$ \\
\hline
\hline
\end{tabular}
\caption{\label{tab:parameters} Quark model parameters.}
\end{center}
\end{table}

Among the different methods to solve the Schr\"odinger equation in order to 
find the quark-antiquark bound states, we use the Gaussian Expansion
Method~\cite{Hiyama:2003cu} which provides enough accuracy and it simplifies the
subsequent evaluation of the decay amplitude matrix elements.

This procedure provides the radial wave function solution of the Schr\"odinger
equation as an expansion in terms of basis functions
\begin{equation}
R_{\alpha}(r)=\sum_{n=1}^{n_{max}} c_{n}^\alpha \phi^G_{nl}(r),
\end{equation} 
where $\alpha$ refers to the channel quantum numbers. The coefficients,
$c_{n}^\alpha$, and the eigenvalue, $E$, are determined from the Rayleigh-Ritz
variational principle
\begin{equation}
\sum_{n=1}^{n_{max}} \left[\left(T_{n'n}^\alpha-EN_{n'n}^\alpha\right)
c_{n}^\alpha+\sum_{\alpha'}
\ V_{n'n}^{\alpha\alpha'}c_{n}^{\alpha'}=0\right],
\end{equation}
where $T_{n'n}^\alpha$, $N_{n'n}^\alpha$ and $V_{n'n}^{\alpha\alpha'}$ are the 
matrix elements of the kinetic energy, the normalization and the potential, 
respectively. $T_{n'n}^\alpha$ and $N_{n'n}^\alpha$ are diagonal, whereas the
mixing between different channels is given by
$V_{n'n}^{\alpha\alpha'}$.

Following Ref.~\cite{Hiyama:2003cu}, we employ Gaussian trial functions with
ranges  in geometric progression. This enables the optimization of ranges
employing a small number of free parameters. Moreover, the geometric
progression is dense at short distances, so that it enables the description of
the dynamics mediated by short range potentials. The fast damping of the
Gaussian tail does not represent an issue, since we can choose the maximal
range much longer than the hadronic size.

The model described above is not able to reproduce the spectrum of the $P$-wave
charmed-strange mesons. The inconsistency with experiment is mainly due to the
fact that the mass splittings between the $D_{s0}^{\ast}(2317)$, $D_{s1}(2460)$
and $D_{s1}(2536)$ mesons are not well reproduced. The same problem appears in
Lattice QCD calculations~\cite{Mohler:2011ke} or other quark
models~\cite{Godfrey:1985xj}.

In order to improve these mass splittings we follow the proposal of
Ref.~\cite{Lakhina:2006fy} and include one-loop corrections to the OGE potential
as derived by Gupta {\it et al.}~\cite{Gupta:1981pd}. This corrections shows
a spin-dependent term which affects only mesons with different flavour quarks.

The net result is a quark-antiquark interaction that can be written as:
\begin{equation}
V(\vec{r}_{ij})=V_{\rm
OGE}(\vec{r}_{ij})+V_{\rm CON}(\vec{r}_{ij})+V_{\rm OGE}^{\rm 1-loop}
(\vec{r}_{ij}),
\end{equation}
where $V_{\rm OGE}$ and $V_{\rm CON}$ were defined before and are treated
non-perturbatively. $V_{\rm OGE}^{\rm 1-loop}$ is the one-loop correction to OGE
potential which is treated perturbatively. As in the case of $V_{\rm OGE}$ and
$V_{\rm CON}$, $V_{\rm OGE}^{\rm 1-loop}$ contains central, tensor and
spin-orbit contributions given by~\cite{Lakhina:2006fy}
\begin{widetext}
\begin{equation}
\begin{split}
&
V_{OGE}^{\rm 1-loop,C}(\vec{r}_{ij})=0, \\
&
\begin{split}
V_{OGE}^{\rm 1-loop,T}(\vec{r}_{ij}) = \frac{C_{F}}{4\pi}
\frac{\alpha_{s}^{2}}{m_{i}m_{j}}\frac{1}{r^{3}}S_{ij} 
&
\left[\frac{b_{0}}{2}\left(\ln(\mu
r_{ij})+\gamma_{E}-\frac{4}{3}\right)+\frac{5}{12}b_ {0}-\frac{2}{3}C_{A}
\right. \\ & \left.
+\frac{1}{2}\left(C_{A}+2C_{F}-2C_{A}\left(\ln(\sqrt{m_{i}m_{j}}\,r_{ij}
)+\gamma_ {
E}-\frac{4}{3}\right)\right)\right],
\nonumber
\end{split}
\end{split}
\end{equation}
\begin{equation}
\begin{split}
&V_{OGE}^{\rm 1-loop,SO}(\vec{r}_{ij})=\frac{C_{F}}{4\pi}
\frac{\alpha_{s}^{2}}{m_{i}^{2}m_{j}^{2}}\frac{1}{r^{3}}\times \\
&
\begin{split}
\times\Bigg\lbrace (\vec{S}_{+}\cdot\vec{L}) & \Big[
\left((m_{i}+m_{j})^{2}+2m_{i}m_{j}\right)\left(C_{F}+C_{A}-C_{A}
\left(\ln(\sqrt{m_{i}m_{j}}\,r_{ij})+\gamma_{E}\right)\right) \\
&
+4m_{i}m_{j}\left(\frac{b_{0}}{2}\left(\ln(\mu
r_{ij})+\gamma_{E}\right)-\frac{1}{12}b_{0}-\frac{1}{2}C_{F}-\frac{7}{6}C_{A}
+\frac{C_{A}}{2}\left(\ln(\sqrt{m_{i}m_{j}}\,r_{ij})+\gamma_{E}\right)\right)
\\
&
+\frac{1}{2}(m_{j}^{2}-m_{i}^{2})C_{A}\ln\left(\frac{m_{j}}{m_{i}}\right)\Big] 
\end{split} \\
&
\begin{split}
\,\,\,\,\,\,\,+(\vec{S}_{-}\cdot\vec{L}) &
\Big[(m_{j}^{2}-m_{i}^{2})\left(C_{F}+C_{A}-C_{A}\left(\ln(\sqrt{m_{i}m_{j}}\,r_
{ij})+\gamma_{E}\right)\right) \\
&
+\frac{1}{2}(m_{i}+m_{j})^{2}C_{A}\ln\left(\frac{m_{j}}{m_{i}}\right)\Big]
\Bigg\rbrace,
\end{split}
\end{split}
\end{equation}
\end{widetext}
where $C_{F}=4/3$, $C_{A}=3$, $b_{0}=9$, $\gamma_{E}=0.5772$ and the scale
$\mu\sim1\,{\rm GeV}$.


\subsection{The $\mathbf{^{3}P_{0}}$ Decay model}
\label{subsec:3P0model}

Meson strong decay is a complex nonperturbative process that has not yet been
described from first principles of QCD. Several phenomenological models have
been developed to deal with this topic. The most popular is the $^{3}P_{0}$
model~\cite{Micu:1968mk,PhysRevD.8.2223,PhysRevD.9.1415} which assumes that a
quark-antiquark pair is created with vacuum quantum numbers, $J^{PC}=0^{++}$.

An important characteristic, apart from its simplicity, is that the model
provides the gross features of various transitions with only one parameter, the
strength $\gamma$ of the decay interaction. Some attempts have been done to
find possible dependences of the vertex parameter $\gamma$,
see~\cite{Ferretti:2013vua} and references therein. In
Ref.~\cite{Segovia:2012cd} we performed a global fit to the decay widths of the
mesons which belong to charmed, charmed-strange, hidden charm and hidden bottom
sectors and elucidated the dependence on the mass scale of the $^{3}P_{0}$ free
parameter $\gamma$. Further details about the global fit can be found in
Ref.~\cite{Segovia:2012cd}. The running of the strength $\gamma$ of the
$^{3}P_{0}$ decay model is given by
\begin{equation}
\gamma(\mu) = \frac{\gamma_{0}}{\log\left(\frac{\mu}{\mu_{0}}\right)},
\label{eq:fitgamma}
\end{equation}
where $\mu$ is the reduced mass of the quark-antiquark in the decaying meson
and, $\gamma_{0}=0.81\pm0.02$ and $\mu_{0}=(49.84\pm2.58)\,{\rm MeV}$ are
parameters determined by the global fit.

We get a quite reasonable global description of the total decay widths in all
meson sectors, from light to heavy. All the wave functions for the mesons
involved in the open-flavour strong decays are the solutions of the
Schr\"odinger equation with the potential model described above and using the
Gaussian Expansion Method~\cite{Hiyama:2003cu}. Details of the resulting matrix
elements for different cases are given in Ref.~\cite{SegoviaThesis}, here we
proceed to explain briefly the main ingredients in which the model is based.

\subsubsection{Transition operator}

The interaction Hamiltonian involving Dirac quark fields that describes the
production process is given by
\begin{equation}
H_{I}=\sqrt{3}\,g_{s}\int d^{3}x \, \bar{\psi}(\vec{x})\psi(\vec{x}),
\label{eq:IH3P0}
\end{equation}
where we have introduced for convenience the numerical factor $\sqrt{3}$, which
will be cancelled with the color factor.

If we write the Dirac fields in second quantization and keep only the
contribution of the interaction Hamiltonian which creates a $(\mu\nu)$
quark-antiquark pair, we arrive, after a nonrelativistic reduction, to the
following expression for the transition operator
\begin{equation}
\begin{split}
T =& -\sqrt{3} \, \sum_{\mu,\nu}\int d^{3}\!p_{\mu}d^{3}\!p_{\nu}
\delta^{(3)}(\vec{p}_{\mu}+\vec{p}_{\nu})\frac{g_{s}}{2m_{\mu}}\sqrt{2^{5}\pi}
\,\times \\
&
\times \left[\mathcal{Y}_{1}\left(\frac{\vec{p}_{\mu}-\vec{p}_{\nu}}{2}
\right)\otimes\left(\frac{1}{2}\frac{1}{2}\right)1\right]_{0}a^{\dagger}_{\mu}
(\vec{p}_{\mu})b^{\dagger}_{\nu}(\vec{p}_{\nu}),
\label{eq:Otransition2}
\end{split}
\end{equation}
where $\mu$ $(\nu)$ are the spin, flavour and color quantum numbers of the
created quark (antiquark). The spin of the quark and antiquark is coupled to
one. The ${\cal Y}_{lm}(\vec{p}\,)=p^{l}Y_{lm}(\hat{p})$ is the solid harmonic
defined in function of the spherical harmonic.

As in Ref.~\cite{Ackleh:1996yt}, we fix the relation of $g_{s}$ with the
dimensionless constant giving the strength of the quark-antiquark pair creation
from the vacuum as $\gamma=g_{s}/2m$, being $m$ the mass of the created quark
(antiquark). In this convention, values of the scale-dependent strength $\gamma$
in the different quark sectors following Eq.~(\ref{eq:fitgamma}) can be found in
Ref.~\cite{Segovia:2012cd}. We use herein the one corresponding to the
charmed-strange meson sector: $\gamma=0.38$.

\subsubsection{Transition amplitude}

\begin{figure}[!t]
\begin{center}
\epsfig{figure=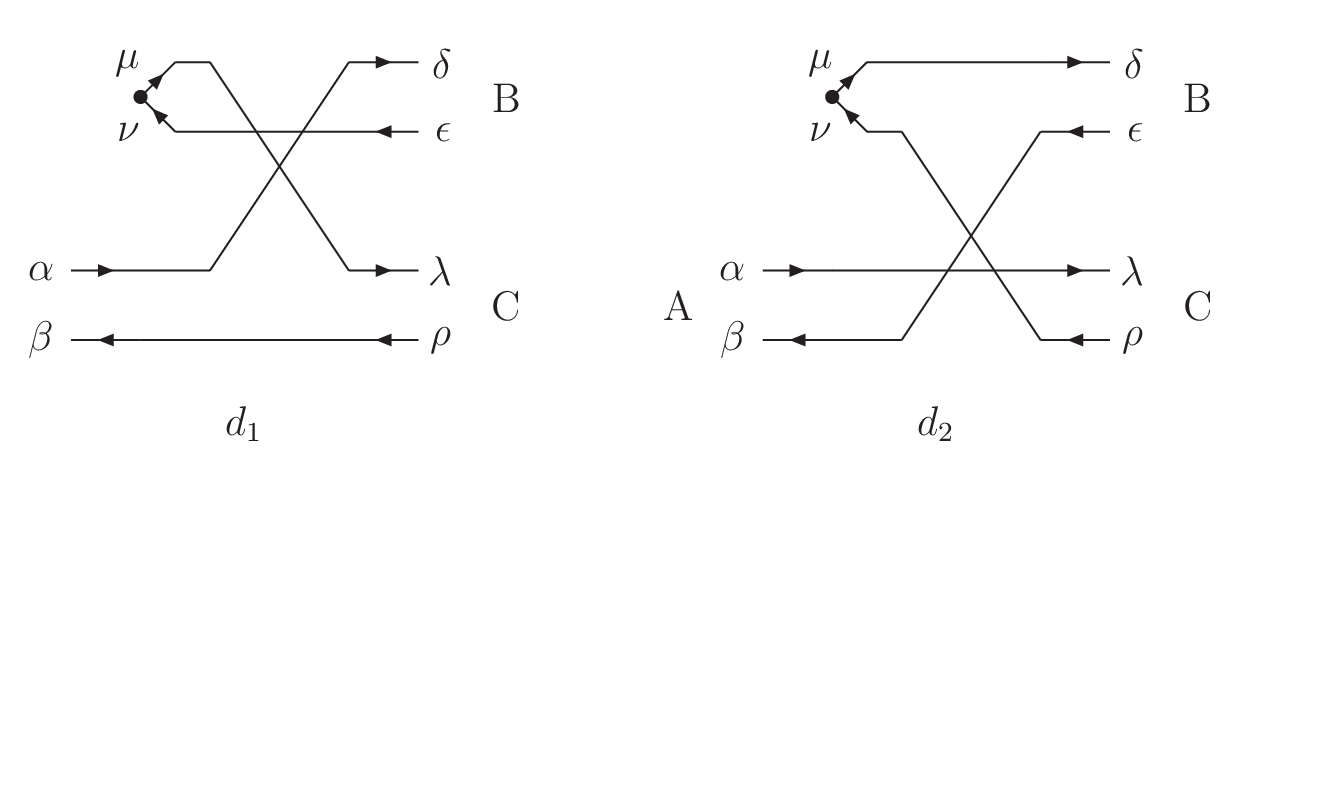,height=3.75cm,width=8.5cm}
\caption{\label{fig:3P0diagrams} Diagrams that can contribute to the decay
width through the $^{3}P_{0}$ model.}
\end{center}
\end{figure}

We are interested on the transition amplitude for the reaction
$(\alpha\beta)_{A} \to (\delta\epsilon)_{B} + (\lambda\rho)_{C}$. The meson $A$
is formed by a quark $\alpha$ and antiquark $\beta$. At some point it is created
a $(\mu\nu)$ quark-antiquark pair. The created $(\mu\nu)$ pair together with the
$(\alpha\beta)$ pair in the original meson regroups in the two outgoing mesons
via a quark rearrangement process. These final mesons are meson $B$ which is
formed by the quark-antiquark pair $(\delta\epsilon)$ and meson $C$ with
$(\lambda\rho)$ quark-antiquark pair.

We work in the center-of-mass reference system of meson $A$, thus we have
$\vec{K}_{A}=\vec{K}_{0}=0$ with $\vec{K}_{A}$ and $\vec{K}_{0}$ the total
momentum of meson $A$ and of the system $BC$ with respect to a given reference
system. We can factorize the matrix element as follow
\begin{equation}
\left\langle BC|T|A\right\rangle =
\delta^{(3)}(\vec{K}_{0}) \mathcal{M}_{A\rightarrow BC}.
\end{equation}

The initial state in second quantization is
\begin{equation}
\left.|A\right\rangle
=\int d^{3}p_{\alpha}d^{3}p_{\beta}\delta^{(3)}(\vec{K}_{A}-\vec{P}_{A})
\phi_{A}(\vec{p}_{A})a_{\alpha}^{\dagger}(\vec{p}_{\alpha})b_{\beta}^{\dagger}
(\vec{p}_{\beta})\left.|0\right\rangle,
\label{eq:Istate}
\end{equation}
where $\alpha$ $(\beta)$ are the spin, flavour and color quantum numbers of
the quark (antiquark). The wave function $\phi_{A}(\vec{p}_{A})$ denotes a meson
$A$ in a color singlet with an isospin $I_{A}$ with projection $M_{I_{A}}$, a
total angular momentum $J_{A}$ with projection $M_{A}$, $J_{A}$ is the coupling
of angular momentum $L_{A}$ and spin $S_{A}$. The $\vec{p}_{\alpha}$ and
$\vec{p}_{\beta}$ are the momentum of quark and antiquark, respectively. The
$\vec{P}_{A}$ and $\vec{p}_{A}$ are the total and relative momentum of the
$(\alpha\beta)$ quark-antiquark pair within the meson $A$. The final state is
more complicated than the initial one because it is a two-meson state. It can be
written as
\begin{widetext}
\begin{equation}
\begin{split}
|BC\!\!\left.\right\rangle =& \frac{1}{\sqrt{1+\delta_{BC}}}\int d^{3}K_{B}
d^{3}K_{C}\sum_{m,M_{BC}}\left\langle\right.
\!\!J_{BC}M_{BC}lm|J_{T}M_{T}\!\!\left.\right\rangle\delta^{(3)}(\vec{K}-\vec{K_
{0}})\delta(k-k_{0}) \\ 
& 
\frac{Y_{lm}(\hat{k})}{k}\sum_{M_{B},M_{C},M_{I_{B}},M_{I_{C}}}\left\langle
J_{B}M_{B}J_{C}M_{C}|J_{BC}M_{BC}\right\rangle\left\langle
I_{B}M_{I_{B}}I_{C}M_{I_{C}}|I_{A}M_{I_{A}} \right\rangle \\ 
& 
\int d^{3}p_{\delta}d^{3}p_{\epsilon}d^{3}p_{\lambda}d^{3}p_{\rho}
\delta^{(3)}(\vec{K}_{B}-\vec{P}_{B})\delta^{(3)}(\vec{K}_{C}-\vec{P}_{C}) \\ 
&
\phi_{B}(\vec{p}_{B})\phi_{C}(\vec{p}_{C})a_{\delta}^{\dagger}(\vec{p}_{\delta}
)b_{\epsilon}^{\dagger}(\vec{p}_{\epsilon})a_{\lambda}^{\dagger}(\vec{p}_{
\lambda})b_{\rho}^{\dagger}(\vec{p}_{\rho})\left.|0\right\rangle,
\label{eq:Fstate}
\end{split}
\end{equation}
\end{widetext}
where we have followed the notation of meson $A$ for the mesons $B$ and $C$. We
assume that the final state of mesons $B$ and $C$ is a spherical wave with
angular momentum $l$. The relative and total momentum of mesons $B$ and $C$ are
$\vec{k}_{0}$ and $\vec{K}_{0}$. The total spin $J_{BC}$ is obtained coupling
the total angular momentum of mesons $B$ and $C$, and $J_{T}$ is the coupling of
$J_{BC}$ and $l$.

The $^{3}P_{0}$ model takes into account only diagrams in which the $(\mu\nu)$
quark-antiquark pair separates into different final mesons. This was originally
motivated by the experiment and it is known as the Okubo-Zweig-Iizuka
(OZI)-rule~\cite{okubo1963,zweigcern2,iizuka1966systematics} which tells us
that the disconnected diagrams are more suppressed than the connected ones. The
diagrams that can contribute to the decay width through the $^{3}P_{0}$ model
are shown in Fig.~\ref{fig:3P0diagrams}.

\begin{table*}[t!]
\begin{center}
\begin{tabular}{ccccccccccccc}
\hline
\hline
\tstrut
& $D_{s}$ & & $D_{s}^{\ast}$ & & $D_{s0}^{\ast}(2317)$ & & $D_{s1}(2460)$ & &
$D_{s1}(2536)$ & & $D_{s2}(2573)$ & \\
\hline
This work $(\alpha_{s})$ & $1984$ & & $2110$ & & $2510$ & & $2593$ & & $2554$ &
& $2591$ & \\
This work $(\alpha_{s}^{2})$ & $1984$ & & $2104$ & & $2383$ & & $2570$ & &
$2560$ & & $2609$ & \\
Experiment & $1969.0\pm1.4$ & & $2112.3\pm0.5$ & & $2318.0\pm1.0$ & &
$2459.6\pm0.9$ & & $2535.18\pm0.24$ & & $2571.9\pm0.8$ & \\
\hline
\hline
\end{tabular}
\caption{\label{tab:1loopDmesons} Masses, in MeV, of the low-lying
charmed-strange mesons predicted by the constituent quark model $(\alpha_{s})$
and those including one-loop corrections to the OGE potential
$(\alpha_{s}^{2})$. Experimental data are taken from
Ref.~\cite{Beringer:1900zz}.}
\end{center}
\end{table*}

\subsubsection{Decay width}

The total width is the sum over the partial widths characterized by the quantum 
numbers $J_{BC}$ and $l$
\begin{equation}
\Gamma_{A\rightarrow BC}=\sum_{J_{BC},l}\Gamma_{A\rightarrow BC}(J_{BC},l),
\end{equation}
where
\begin{equation}
\Gamma_{A\rightarrow BC}(J_{BC},l)=2\pi\int
dk_{0}\delta(E_{A}-E_{BC})|\mathcal{M}_{A\rightarrow BC}(k_{0})|^{2}.
\label{eq:gammaDelta}
\end{equation}

We use relativistic phase space, so
\begin{equation}
\begin{split}
\Gamma_{A\rightarrow 
BC}(J_{BC},l)=2\pi\frac{E_{B}(k_{0})E_{C}(k_{0})}{m_{A}k_{0}}|\mathcal{M}_{
A\rightarrow BC}(k_{0})|^{2},
\end{split}
\end{equation}
where
\begin{equation}
k_{0}=\frac{\sqrt{[m_{A}^{2}-(m_{B}-m_{C})^{2}][m_{A}^{2}-(m_{B}+m_{C})^{2}]}}{
2m_{A}},
\end{equation}
is the on-shell relative momentum of mesons $B$ and $C$.


\section{RESULTS}
\label{sec:results}

\subsection{Review of low-lying states}

Table~\ref{tab:1loopDmesons} shows the masses of the low-lying charmed-strange
mesons predicted by the constituent quark model. One can see our results
taking into account the one-gluon exchange potential $(\alpha_{s})$ and
including its one-loop corrections $(\alpha_{s}^{2})$.

The theoretical masses for the $D_{s}$ and $D_{s}^{\ast}$ mesons agree with the
experimental measurements. We also find a reasonable agreement for the
$D_{s1}(2536)$ and $D_{s2}(2573)$ masses. The state assigned to the
$D_{s0}^{\ast}(2317)$ is very sensitive to the one-loop corrections of the OGE
potential which bring its mass closer to the experimental one. This effect could
explain part of its lower mass as a $c\bar{s}$ state, but threshold effects
should be taken into account before a definitive statement can be given about
its nature. The spin dependent corrections to the OGE potential are not enough
to solve the puzzle in the $1^{+}$ $c\bar{s}$ sector. In
Ref.~\cite{Segovia:2009zz} we have studied the $J^{P}=1^{+}$ charmed-strange
channel, finding that the $D_{s1}(2460)$ has an important non-$q\bar q$
contribution whereas the $D_{s1}(2536)$ is almost a pure $q\bar{q}$ state.
However, the presence of non-$q\bar{q}$ degrees of freedom modifies the
$D_{s1}(2536)$ wave function in such away that explains most of its decay
properties~\cite{Segovia:2009zz,Segovia:2011dg,Segovia:2012yh}. 

\begin{table}[!t]
\begin{center}
\begin{tabular}{lllccc}
\hline
\hline
\tstrut
Meson & $n\,J^{P}$ & Channel & $\Gamma_{^{3}P_{0}}$ & ${\cal B}_{^{3}P_{0}}$ &
$\Gamma_{\rm exp.}$ \\
& & & (MeV) & (\%) & (MeV) \\
\hline
\tstrut
$D_{s1}(2536)^{+}$ & $1\,1^{+}$ & $D^{\ast+}K^{0}$ & $0.43$ & $43.48$ & \\
& & $D^{\ast0}K^{+}$ & $0.56$ & $56.52$ & \\
& & total & $0.99$ & $100$ & $0.92\pm0.05$ \\
\hline
\tstrut
$D_{s2}(2573)^{+}$ & $1\,2^{+}$ & $D^{+}K^{0}$ & $8.02$ & $42.95$ & \\
& & $D^{0}K^{+}$ & $8.69$ & $46.54$ & \\
& & $D^{\ast+}K^{0}$ & $0.82$ & $4.40$ & \\
& & $D^{\ast0}K^{+}$ & $1.06$ & $5.67$ & \\
& & $D_{s}^{+}\eta$ & $0.08$ & $0.44$ & \\
& & total & $18.67$ & $100$ & $17\pm4$ \\
\hline
\hline
\end{tabular}
\caption {\label{tab:lowerDs} Open-flavour strong decay widths, in MeV, and
branching fractions, in $\%$, of the $D_{s1}(2536)$ and $D_{s2}^{\ast}(2573)$
mesons. Experimental data are taken from Ref.~\cite{Beringer:1900zz}.}
\end{center}
\end{table}

Table~\ref{tab:lowerDs} shows the partial and total strong decay widths of the
mesons $D_{s1}(2536)$ and $D_{s2}^{\ast}(2573)$. We show the absolute values in
MeV and the branching fractions in \%. One can see that the total decay widths
reported by PDG~\cite{Beringer:1900zz} are in excellent agreement with our
results. For the $D_{s1}(2536)$ meson, the PDG provides the following two-body
decay branching ratios (concerning strong decays):
\begin{equation}
\begin{split}
R_{1} &= \frac{\Gamma(D_{s1}(2536)^{+} \rightarrow D^{\ast0}K^{+})}
{\Gamma(D_{s1}(2536)^{+} \rightarrow D^{\ast+}K^{0})} = 1.18\pm0.16, \\
R_{2} &= \frac{\Gamma_{S}(D_{s1}(2536)^{+} \rightarrow D^{\ast+}K^{0})}
{\Gamma(D_{s1}(2536)^{+} \rightarrow D^{\ast+}K^{0})} = 0.72\pm0.05\pm0.01,
\end{split}
\end{equation}
which compare reasonably well with our theoretical results for $R_{1}=1.31$ and
$R_{2}=0.66$. For $D_{s1}^{\ast}(2573)$ meson, the PDG only reports an upper
limit on the ratio $D^{\ast0}K^{+}/D^{0}K^{+}$ of $0.33$. Our theoretical figure
$0.12$ is compatible with such a limit. It is worth to mention here that in the
work in which the LHCb Collaboration disentangles the resonance structure of the
peak around $2.86\,{\rm GeV}$~\cite{Aaij:2014baa}, they also provide
the $D_{s2}^{\ast}(2573)$ mass and width with significantly better precision
than previous measurements
\begin{equation}
\begin{split}
M(D_{s2}(2573)) &= (2568.39 \pm 0.29 \pm 0.19 \pm 0.18) \, {\rm MeV}, \\ 
\Gamma(D_{s2}(2573)) &= (16.9 \pm 0.5 \pm 0.4 \pm 0.4) \, {\rm MeV},
\end{split}
\end{equation}
and both are in reasonable agreement with our theoretical results: $2.61\,{\rm
GeV}$ and $18.67\,{\rm MeV}$, respectively.


\subsection{The $\mathbf{D_{s1}^{\ast}(2700)}$ resonance}
\label{subsec:Ds1-2700}

It is commonly believed that the $D_{s1}^{\ast}(2700)$ is the first excitation
of the $D_{s}^{\ast}$ meson. Our quark model predicts a mass in this energy
range ($2.79\,{\rm GeV}$) but also for the $n^{2S+1}L_{J}=2^{1}S_{0}$ state
($2.73\,{\rm GeV}$). However, if the $D_{s1}^{\ast}(2700)$ had quantum numbers
$J^{P}=0^{-}$ it would not decay into $DK$ final state, and this is incompatible
with the experimental observations. Table~\ref{tab:Ds2700} shows the
open-flavour strong decays of the $D_{s1}^{\ast}(2700)$ meson as the
$2^{3}S_{1}$ state.

\begin{table}[!t]
\begin{center}
\begin{tabular}{lllccc}
\hline
\hline
\tstrut
Meson & $n\,J^{P}$ & Channel & $\Gamma_{^{3}P_{0}}$ & ${\cal B}_{^{3}P_{0}}$ &
$\Gamma_{\rm exp.}$ \\
& & & (MeV) & (\%) & (MeV) \\
\hline
\tstrut
$D_{s1}^{\ast}(2700)$  & $2\,1^{-}$ & $DK$ & $36.99$ & $21.67$ & \\
& & $D^{\ast}K$            & $97.78$  & $57.26$ & \\
& & $D_{s}\eta$            & $3.67$   & $2.15$ & \\
& & $D_{s}^{\ast}\eta$     & $9.51$   & $5.57$ & \\
& & $D^{\ast}K_{0}^{\ast}$ & $22.80$  & $13.35$ & \\
& & total                  & $170.75$ & $100$ & $125\pm30$ \\
\hline
\hline
\end{tabular}
\caption {\label{tab:Ds2700} Open-flavour strong decay widths, in MeV, and
branching fractions, in $\%$, of the $D_{s1}^{\ast}(2700)$ meson with quantum
numbers $nJ^{P}=2\,1^{-}$. Experimental data are taken from
Ref.~\cite{Beringer:1900zz}.}
\end{center}
\end{table}

The total decay width of $D_{s1}^{\ast}(2700)$ as the $2^{3}S_{1}$ state is
slightly larger but close to the experimental value. The information of
the partial decay widths shown in Table~\ref{tab:Ds2700} points out that the
$D^{\ast}K$ decay channel is dominant and the $DK$ and $D^{\ast}K_{0}^{\ast}$
are important. In addition, the $D_{s1}^{\ast}(2700)$ meson has traces in
$D_{s}\eta$ and $D_{s}^{\ast}\eta$ with partial widths of several MeV. Finally,
our theoretical value for the branching ratio $D^{\ast}K/DK$ is $2.6$, which is
a factor $3$ larger than the experimental measurement $0.91\pm0.13\pm0.12$
reported by the BaBar Collaboration~\cite{Aubert:2009ah}. Similar discrepancies
can be found in other quark models. This fact may be an indication of a bigger
mixture between the $2^{3}S_{1}$ and $1^{3}D_{1}$ states~\cite{Song:2015nia}. In
our model the mixing is not fitted to the experimental data but driven by the
tensor piece of the quark-antiquark interaction. Our states are almost pure
$^3S_1$ or $^3D_1$ in the $J^{P}=1^{-}$ channel. It would be very helpful that
the LHCb Collaboration, with significantly better precision than previous
experiments, repeats the measurement of this ratio.


\subsection{The $\mathbf{D_{sJ}^{\ast}(2860)}$ resonance}
\label{subsec:DsJ-12860}

According to the observed decay modes, the possible spin-parity quantum numbers
of the $D_{sJ}^{\ast}(2860)$ are $J^{P}=1^{-}$, $2^{+}$, $3^{-}$, and so on. The
$2^{+}$ assignment is disfavoured because it would be the excitation of the
$D_{s2}^{\ast}(2573)$ meson and our model predicts a mass around $3.1\,{\rm
GeV}$. Beyond $J=3$ the predicted masses are much higher than the
experimental measurement. Table~\ref{tab:Ds2860} shows the open-flavour strong
decays of the $D_{sJ}^{\ast}(2860)$ as the third excitation of the $1^{-}$ meson
(mostly dominated by the $1^{3}D_{1}$ channel) and as the ground state of
$3^{-}$ meson (mostly dominated by the $1^{3}D_{3}$ channel).

\begin{table}[!t]
\begin{center}
\begin{tabular}{lllccc}
\hline
\hline
Meson & $n\,J^{P}$ & Channel & $\Gamma_{^{3}P_{0}}$ & ${\mathcal B}_{^3P_0}$ 
& $\Gamma_{\rm exp.}$ \\
& & & (MeV) & (\%) & (MeV) \\
\hline
$D_{sJ}^{\ast}(2860)$ & $3\,1^{-}$ & $DK$ & $53.34$ & $34.81$ & \\
& & $D^{\ast}K$            & $38.43$  & $25.08$ & \\
& & $D_{s}\eta$            & $12.12$  & $7.92$  & \\
& & $D_{s}^{\ast}\eta$     & $5.06$   & $3.30$  & \\
& & $D^{\ast}K_{0}^{\ast}$ & $7.15$   & $4.67$  & \\
& & $DK^{\ast}$            & $37.10$  & $24.22$ & \\
& & total                  & $153.20$ & $100$   & $159\pm23\pm27\pm72$ \\
\hline
$D_{sJ}^{\ast}(2860)$ & $1\,3^{-}$ & $DK$ & $38.57$ & $45.32$ & \\
& & $D^{\ast}K$            & $26.17$  & $30.74$ & \\
& & $D_{s}\eta$            & $1.06$   & $1.24$  & \\
& & $D_{s}^{\ast}\eta$     & $0.35$   & $0.41$  & \\
& & $D^{\ast}K_{0}^{\ast}$ & $16.16$  & $18.99$ & \\
& & $DK^{\ast}$            & $2.81$   & $3.30$  & \\
& & total                  & $85.12$  & $100$   & $53\pm7\pm4\pm6$ \\
\hline
\hline
\end{tabular}
\caption {\label{tab:Ds2860} Open-flavour strong decay widths, in MeV, and
branching fractions, in $\%$, of the $D_{sJ}^{\ast}(2860)$ meson with quantum
numbers $nJ^{P}=3\,1^{-} \mbox{ or }1\,3^{-}$. Experimental data are taken
from Refs.~\cite{Aaij:2014xza,Aaij:2014baa}}
\end{center}
\end{table}

The predicted total decay widths are in excellent agreement with the LHCb
observation of having two resonances at $2.86\,{\rm GeV}$, one of spin-$1$
and another one of spin-$3$. Table~\ref{tab:Ds2860} shows that the
$DK$, $D^{\ast}K$ and $DK^{\ast}$ decay channels are very important for the
$D_{s1}^{\ast}(2860)$ meson. However, the $D_{s3}^{\ast}(2860)$ resonance
decays mainly into $DK$, $D^{\ast}K$ and $D^{\ast}K^{\ast}_{0}$ final states
being the partial width into $DK^{\ast}$ very small. We also observe that
the $D_{s1}^{\ast}(2860)$ meson has traces in $D_{s}\eta$ and $D_{s}^{\ast}\eta$
while the partial decay widths of the $D_{s3}^{\ast}(2860)$ into these final
states are very tiny. With respect the branching ratio measured by the
BaBar Collaboration~\cite{Aubert:2009ah}, we obtain
\begin{equation}
\frac{{\cal{B}}(D_{sJ}^{\ast}(2860) \to D^{\ast}K)}
{{\cal{B}}(D_{sJ}^{\ast}(2860) \to DK)} = \begin{cases} 0.72 &
D_{s1}^{\ast}(2860), \\ 0.68 & D_{s3}^{\ast}(2860), \end{cases}
\end{equation}
which compares reasonably well with the experimental one, $1.10\pm0.15\pm0.19$.
In view of our results, we cannot distinguish if the branching ratio measured by
BaBar belongs to the $D_{s1}^{\ast}(2860)$ or to the $D_{s3}^{\ast}(2860)$. It
would be again very helpful that the LHCb Collaboration repeats the measurement
of this ratio.


\subsection{The $\mathbf{D_{sJ}(3040)}$ resonance}
\label{subsec:DsJ-3040}

\begin{table}[!t]
\begin{center}
\begin{tabular}{lllccc}
\hline
\hline
Meson & $n\,J^{P}$ & Channel & $\Gamma_{^{3}P_{0}}$ & ${\mathcal B}_{^3P_0}$ &
$\Gamma_{\rm exp.}$ \\
\hline
$D_{sJ}(3040)$ & $nJ^{P}=3\,1^{+}$ & $D^{\ast}K$ & $25.22$ & $8.36$ & \\
& & $DK_{0}^{\ast}$        & $0.76$   & $0.25$  & \\
& & $D_{s}^{\ast}\eta$     & $3.26$   & $1.08$  & \\
& & $D^{\ast}K_{0}^{\ast}$ & $0.02$   & $0.01$  & \\
& & $DK^{\ast}$            & $44.28$  & $14.69$ & \\
& & $D_{s0}^{\ast}\eta$    & $0.97$   & $0.32$  & \\
& & $D^{\ast}_{0}K$        & $2.81$   & $0.93$  & \\
& & $D^{\ast}K^{\ast}$     & $156.78$ & $52.00$ & \\
& & $D_{1}K$               & $39.81$  & $13.20$ & \\
& & $D'_{1}K$              & $0.69$   & $0.23$  & \\
& & $D_{2}^{\ast}K$        & $11.19$  & $3.71$  & \\
& & $D_{s}\phi$            & $15.54$  & $5.15$  & \\
& & $D_{s1}(2460)\eta$     & $0.19$   & $0.07$  & \\
& & total                  & $301.52$ & $100$   & $239 \pm 35^{+46}_{-42}$ \\
\hline
$D_{sJ}(3040)$ & $nJ^{P}=4\,1^{+}$ & $D^{\ast}K$ & $53.48$ & $12.37$ & \\
& & $DK_{0}^{\ast}$        & $0.30$   & $0.07$  & \\
& & $D_{s}^{\ast}\eta$     & $4.97$   & $1.15$  & \\
& & $D^{\ast}K_{0}^{\ast}$ & $1.10$   & $0.25$  & \\
& & $DK^{\ast}$            & $100.38$ & $23.21$ & \\
& & $D_{s0}^{\ast}\eta$    & $1.66$   & $0.38$  & \\
& & $D^{\ast}_{0}K$        & $2.31$   & $0.53$  & \\
& & $D^{\ast}K^{\ast}$     & $130.91$ & $30.27$ & \\
& & $D_{1}K$               & $11.58$  & $2.68$  & \\
& & $D'_{1}K$              & $0.04$   & $0.01$  & \\
& & $D_{2}^{\ast}K$        & $123.74$ & $28.61$ & \\
& & $D_{s}\phi$            & $1.97$   & $0.45$  & \\
& & $D_{s1}(2460)\eta$     & $0.09$   & $0.02$  & \\
& & total                  & $432.53$ & $100$   & $239 \pm 35^{+46}_{-42}$ \\
\hline
\hline
\end{tabular}
\caption {\label{tab:Ds3040} Open-flavour strong decay widths, in MeV, and
branching fractions, in $\%$, of the $D_{sJ}(3040)$ meson with quantum numbers
$nJ^{P}=3\,1^{+} \mbox{or } 4\,1^{+}$. Experimental data are taken from
Ref.~\cite{Beringer:1900zz}.}
\end{center}
\end{table}

\begin{table}[!t]
\begin{center}
\begin{tabular}{lllcc}
\hline
\hline
Meson & $n\,J^{P}$ & Channel & $\Gamma_{^{3}P_{0}}$ & ${\mathcal B}_{^3P_0}$ \\
\hline
$D_{s0}^{\ast}(2934)$ & $2\,0^{+}$ & $DK$ & $60.62$ & $32.67$ \\
& & $D_{s}\eta$                & $3.31$   & $1.78$ \\
& & $D^{\ast}K_{0}^{\ast}$     & $4.47$   & $2.41$ \\
& & $D^{\ast}K^{\ast}$         & $86.42$  & $46.57$ \\
& & $D_{1}K$                   & $18.43$  & $9.93$ \\
& & $D'_{1}K$                  & $0.14$   & $0.08$ \\
& & $D_{s}\eta'$               & $12.17$  & $6.56$ \\
& & total                      & $185.56$ & $100$ \\
\hline
$D_{s2}^{\ast}(3094)$ & $2\,2^{+}$ & $DK$ & $2.14$ & $0.95$ \\
& & $D^{\ast}K$                & $1.90$   & $0.84$ \\
& & $D_{s}\eta$                & $0.02$   & $0.01$ \\
& & $D_{s}^{\ast}\eta$         & $1.02$   & $0.45$ \\
& & $D^{\ast}K_{0}^{\ast}$     & $0.80$   & $0.36$ \\
& & $DK^{\ast}$                & $40.32$  & $17.91$ \\
& & $D^{\ast}K^{\ast}$         & $124.26$ & $55.18$ \\
& & $D_{1}K$                   & $17.59$  & $7.81$ \\
& & $D'_{1}K$                  & $6.35$   & $2.82$ \\
& & $D_{s}\eta'$               & $1.71$   & $0.76$ \\
& & $D_{2}^{\ast}K$            & $26.30$  & $11.68$ \\
& & $D_{s}\phi$                & $0.00$   & $0.00$ \\
& & $D_{s1}(2460)\eta$         & $2.44$   & $1.09$ \\
& & $D_{0}^{\ast}K_{0}^{\ast}$ & $0.32$   & $0.14$ \\
& & total                      & $225.17$ & $100$ \\
\hline
\hline
\end{tabular}
\caption {\label{tab:missing1} Open-flavour strong decay widths, in MeV, and
branching fractions, in $\%$, of the $nJ^{P}=2\,0^{+}$ and $nJ^{P}=2\,2^{+}$
states.}
\end{center}
\end{table}

The mean $2P$ multiplet mass is predicted in our model to be $3.06\,{\rm GeV}$
which is near the mass of the $D_{sJ}(3040)$ resonance. Therefore, the possible
assignments are the $J^{P}=0^{+}$ which only decays into $DK$, the $1^{+}$ which
only decays into $D^{\ast}K$ and the $2^{+}$ which decays into $DK$ and
$D^{\ast}K$. The only decay mode in which $D_{sJ}(3040)$ has been seen until now
is the $D^{\ast}K$, and so the most possible assignment is that the
$D_{sJ}(3040)$ meson being the next excitation in the $1^{+}$ channel. 

Table~\ref{tab:Ds3040} shows the open-flavour strong decays of the
$D_{sJ}(3040)$ meson as the $n\,J^{P}=3\,1^{+}$ or $4\,1^{+}$ state. The mass of
the $D_{sJ}(3040)$ is large enough to allow open-flavour strong decays not
studied before in this work like the $D^{\ast}K^{\ast}$ final state. In fact,
this decay channel is dominant in both $n\,J^{P}=3\,1^{+}$ and $4\,1^{+}$ states
followed by the $DK^{\ast}$ in the case of the $3\,1^{+}$ and by the
$D_{2}^{\ast}K$ and $DK^{\ast}$ in the case of the $4\,1^{+}$. Moreover, the
$n\,J^{P}=3\,1^{+}$ state has partial decay widths in the order of tens of MeV
for the $D^{\ast}K$, $D_{1}K$, $D_{s}\phi$ and $D_{2}^{\ast}K$ decay channels;
whereas for the $nJ^{P}=4\,1^{+}$ we find only partial widths in the order of
tens of MeV for the $D^{\ast}K$ and $D_{1}K$. Finally, the total decay widths
are large for both states, being that of the $nJ^{P}=3\,1^{+}$ state in better
agreement with the experimental data.

There are other two states which belong to the mean $2P$ multiplet that are
still missing in experiment. These states are the first radial excitation of the
$c\bar{s}$ mesons with quantum numbers $J^{P}=0^{+}$ and $2^{+}$, respectively.
The masses predicted by our model are $2.93\,{\rm GeV}$ for the
$nJ^{P}=2\,0^{+}$ and $3.09\,{\rm GeV}$ for the $2\,2^{+}$. These masses are in
agreement with recent studies of the charmed-strange meson
sector~\cite{Godfrey:2014fga,Song:2015nia}. It is worthy to remind here that the
$J^{P}=0^{+}$ channel is very sensitive to the $1$-loop correction of the OGE
potential. For the second excitation, its mass goes from $3.03$ to $2.93\,{\rm
GeV}$.

Table~\ref{tab:missing1} shows their partial and total decay widths calculated
with the $^{3}P_{0}$ decay model. The masses used for the initial mesons are
the theoretical ones. There are less open decay channels for the
$nJ^{P}=2\,0^{+}$ resonance than for the $2\,2^{+}$. However, the total decay
widths are very similar being $185.55\,{\rm MeV}$ and $225.17\,{\rm MeV}$,
respectively. The $D^{\ast}K^{\ast}$ decay channel is dominant for the two
resonances. One can also find traces of the $D_{1}K$ decay channel in both
cases, but the $DK^{\ast}$ and $D_{2}^{\ast}K$ decays are only important for the
$2\,2^{+}$ resonance. It is remarkably that the $2\,0^{+}$ resonance has large
partial width to $DK$ whereas this is not the case for the $2\,2^{+}$ state. The
authors of Ref.~\cite{Song:2015nia} predict the same behaviour than us and
notice that there is an evidence of a structure around $2.96\,{\rm GeV}$ in the
$\bar{D}^{0}K^{-}$ invariant mass spectrum given by
LHCb~\cite{Aaij:2014xza,Aaij:2014baa} that can be associated to the $2\,0^{+}$
resonance.


\subsection{The missing states around $\mathbf{2.86\,{\rm GeV}}$}

\begin{table}[!t]
\begin{center}
\begin{tabular}{lllcc}
\hline
\hline
Meson & $n\,J^{P}$ & Channel & $\Gamma_{^{3}P_{0}}$ & ${\mathcal B}_{^3P_0}$ \\
\hline
$D_{s}(2729)$ & $2\,0^{-}$ & $D^{\ast}K$ & $190.62$  & $66.81$ \\
& & $DK_{0}^{\ast}$    & $87.41$  & $30.64$ \\
& & $D_{s}^{\ast}\eta$ & $7.28$   & $2.55$  \\
& & total              & $285.31$ & $100$   \\
\hline
$D_{s2}(2888)$ & $1\,2^{-}$ & $D^{\ast}K$ & $54.74$ & $23.45$ \\
& & $DK_{0}^{\ast}$        & $35.92$  & $15.39$ \\
& & $D_{s}^{\ast}\eta$     & $0.99$   & $0.43$  \\
& & $D^{\ast}K_{0}^{\ast}$ & $8.19$   & $3.51$  \\
& & $DK^{\ast}$            & $133.55$ & $57.22$ \\
& & $D_{s0}^{\ast}\eta$    & $0.00$   & $0.00$  \\
& & $D^{\ast}_{0}K$        & $0.00$   & $0.00$  \\
& & total                  & $233.39$ & $100$   \\
\hline
$D_{s2}(2948)$ & $2\,2^{-}$ & $D^{\ast}K$ & $85.70$ & $45.86$ \\
& & $DK_{0}^{\ast}$        & $4.52$   & $2.42$ \\
& & $D_{s}^{\ast}\eta$     & $19.10$  & $10.22$ \\
& & $D^{\ast}K_{0}^{\ast}$ & $9.51$   & $5.09$ \\
& & $DK^{\ast}$            & $31.61$  & $16.92$ \\
& & $D_{s0}^{\ast}\eta$    & $0.00$   & $0.00$ \\
& & $D^{\ast}_{0}K$        & $0.02$   & $0.01$ \\
& & $D^{\ast}K^{\ast}$     & $36.30$  & $19.42$ \\
& & $D_{1}K$               & $0.12$   & $0.06$ \\
& & $D'_{1}K$              & $0.01$   & $0.00$ \\
& & $D_{s}\eta'$           & $0.00$   & $0.00$ \\
& & total                  & $186.89$ & $100$ \\
\hline
\hline
\end{tabular}
\caption {\label{tab:missing2} Open-flavour strong decay widths, in MeV, and
branching fractions, in $\%$, of the $nJ^{P}=2\,0^{-}$, $nJ^{P}=1\,2^{-}$ and 
$nJ^{P}=2\,2^{-}$ states.}
\end{center}
\end{table}

We have assigned to the $D_{s1}^{\ast}(2700)$ meson the $n\,J^{P}=2\,1^{-}$
state and to the $D_{sJ}^{\ast}(2860)$ meson the $3\,1^{-}$ and $1\,3^{-}$
states. The last two assignments are based on the disentanglement of spin-$1$
and spin-$3$ resonances around the $2.86\,{\rm GeV}$ peak performed by the LHCb
Collaboration~\cite{Aaij:2014xza,Aaij:2014baa}. Around this energy range, there
are still three states not seen by experiments. These are the second radial
excitation of the charmed-strange ground state $n\,J^{P}=2\,0^{-}$ and the
$1\,2^{-}$ and $2\,2^{-}$ states which are the $D$-wave partners of the
$D_{sJ}^{\ast}(2860)$ meson. 

The masses predicted by our constituent quark model are $2.79\,{\rm GeV}$,
$2.89\,{\rm GeV}$ and $2.95\,{\rm GeV}$ for the $n\,J^{P}=2\,0^{-}$, $1\,2^{-}$
and $2\,2^{-}$, respectively. These masses are in agreement with
Refs.~\cite{Godfrey:2014fga,Song:2015nia} except for the $2\,0^{-}$ state which
seems that our model predicts a mass slightly higher. Table~\ref{tab:missing2}
shows partial and total decay widths for the three states. One can see that we
predict a $2\,0^{-}$ state much broader than recent studies of the same
states~\cite{Godfrey:2014fga,Song:2015nia}. There are two main reasons for this:
i) our theoretical mass is different and this influences the calculation of the
decay widths in the $^{3}P_{0}$ model; ii)
Refs.~\cite{Godfrey:2014fga,Song:2015nia} do not calculate the partial width
into the $DK_{0}^{\ast}$ despite this decay channel is open for the mass they
predict. The $DK_{0}^{\ast}$ decay channel contributes $30\%$ to the
total decay width in our model and sum almost $100\,{\rm MeV}$ to it.

With respect the other two states, $1\,2^{-}$ and $2\,2^{-}$, their masses and
total decay widths are very similar to those predicted in
Refs.~\cite{Godfrey:2014fga,Song:2015nia}. One can distinguish common
features between their predictions and ours. The first one is that these states
seems quite broad with total decay widths around $200\,{\rm MeV}$. The second
one is that the dominant decay channels for the $2\,2^{-}$ state are
$D^{\ast}K$, $DK^{\ast}$ and $D^{\ast}K^{\ast}$ with traces also in the
$D_{s}^{\ast}\eta$. However, there are also differences. The relative order in
the dominant channels for the $2\,2^{-}$ state is different and, what is more
important, our predictions for the partial decay widths of the
$n\,J^{P}=1\,2^{-}$ seems quite different. They predict that its dominant decay
channel is $D^{\ast}K$ followed by the $D_{s}^{\ast}\eta$ whereas
we have a dominant $DK^{\ast}$ followed by the $D^{\ast}K$ and $DK_{0}^{\ast}$
decay channels.


\section{SUMMARY AND CONCLUSIONS}
\label{sec:conclusions}

We have performed an extensive study of strong decay properties for the
$D_{s1}^{\ast}(2860)$ and $D_{s3}^{\ast}(2860)$ resonances. We have completed
the study  with the same analysis for the $D_{s1}^{\ast}(2700)$ and
$D_{sJ}(3040)$. Our theoretical results indicate that the $D_{s1}^{\ast}(2700)$,
$D_{s1}^{\ast}(2860)$, $D_{s3}^{\ast}(2860)$ and $D_{sJ}(3040)$ mesons can be
accommodated as the $n\,J^{P}=2\,1^{-}$, $3\,1^{-}$, $1\,3^{-}$ and $3\,1^{+}$,
respectively. These predictions are in agreement with other theoretical studies
of the same resonances.

For this study we have used a constituent quark model which describes
successfully the hadron phenomenology from light to heavy quark sectors. The
$1$-loop corrections to the OGE potential and the coupling of non-$q\bar{q}$
degrees of freedom in the $1^{+}$ $c\bar{s}$ channel improve the theoretical
mass splittings between the $D_{s0}^{\ast}(2317)$, $D_{s1}(2460)$ and
$D_{s1}(2536)$ mesons. Moreover, the decay properties of the $D_{s1}(2536)$
and $D_{s2}(2573)$ are well reproduced. It is worth to mention that the strong
decays have been calculated using an adapted version of the $^{3}P_{0}$ decay
model in which the strength $\gamma$ of the decay interaction depends on the
mass scale through the reduced mass of the quark-antiquark in the decaying
meson.

Finally, there are states still undiscovered by experiments in the mass energy
region of the $D_{s1}^{\ast}(2700)$, $D_{sJ}^{\ast}(2860)$ and $D_{sJ}(3040)$
mesons. We have provided the masses and strong decay properties of those which
belong to the same spin-multiplets. We hope that this study will help
experimentalists in carrying out a search for them.

\begin{acknowledgments}
This work has been partially funded by Ministerio de Ciencia y Tecnolog\'\i a
under Contract no. FPA2013-47443-C2-2-P, by the European Community-Research
Infrastructure Integrating Activity ``Study of Strongly Interacting Matter''
(HadronPhysics3 Grant no. 283286) and by the Spanish Ingenio-Consolider 2010
Program CPAN (CSD2007-00042). JS acknowledges financial support from a
postdoctoral IUFFyM contract of the Universidad de Salamanca.
\end{acknowledgments}


\bibliographystyle{apsrev}
\bibliography{Ds2860}

\end{document}